\documentclass[aps,pra,showpacs,twocolumn]{revtex4}
\usepackage{graphicx,color}
\usepackage{amsmath, amssymb}

\begin{document}
\title{Representation of entanglement by negative quasi-probabilities}

\author{J. Sperling} \email{jan.sperling2@uni-rostock.de}
\author{W. Vogel} \email{werner.vogel@uni-rostock.de}
\affiliation{Arbeitsgruppe Quantenoptik, Institut f\"ur Physik, Universit\"at Rostock, D-18051 Rostock, Germany}

\date{\today}

\begin{abstract}
Any bipartite quantum state has quasi-probability representations in terms of separable states.
For entangled states these quasi-probabilities necessarily exhibit negativities. Based on the general structure of composite quantum states, one may reconstruct such quasi-propabilities from experimental data. Because of ambiguity, the quasi-probabilities obtained by the bare reconstruction are insufficient to identify entanglement. 
An optimization procedure is introduced to derive quasi-probabilities with a minimal amount of negativity. Negativities of optimized quasi-probabilities unambiguously prove entanglement, their positivity proves separability.
\end{abstract}

\maketitle

\section{Introduction}

Among the most striking discrepancies between classical and quantum physics are the nonclassical correlations between the subsystems of a compound system~\cite{ERP}.
In general, a common description of all subsystems is needed for describing the properties and the evolution of the compound system~\cite{Schroed}.
These nonclassical correlations are usually called entanglement.
Today, entanglement is considered to be the key resource of the vast fields of Quantum Information Processing, Quantum Computation, and Quantum Technology, see e.g.~\cite{book1,book2}.
The most prominent examples of entangled states are the Bell states, which maximally violate Bell-type inequalities which are valid for classical correlations~\cite{Bell}.

On the other hand, in the field of Quantum Optics nonclassical effects have been studied over many years, which are related to negativities in the Glauber--Sudarshan distribution~\cite{Glauber,Sudar}, in the following denoted as $P_{\rm GS}$.
It describes any quantum state of the harmonic oscillator as a pseudo-mixture of coherent states.
A state is considered to be classical, if the $P_{\rm GS}$ function has the properties of a classical propability distribution.
If this is not the case, then the state is called a nonclassical one~\cite{Titulaer,Mandel}.
Recently, negativities of $P_{\rm GS}$ could be experimentally demonstrated~\cite{Kiesel}.

Entanglement and the nonclassical effects reflected by the negativities in $P_{\rm GS}$ appear to be related in some form.
When mixing two radiation fields by a beam splitter, nonclassicality in one of the input channel is necessary for the generation of entanglement in the output channels~\cite{Kim,Xiang}.
On the other hand, even an entangled composite state has negativities in the $P_{\rm GS}$ function, which is a necessary but not sufficient condition for entanglement.

A detailed understanding of the structure of entangled quantum states is a topic of fundamental interest.
The idea of the decomposition of entangled states into a best approximation by a separable state and a remaining entangled state delivered further insights into the nature of entanglement, see~\cite{Karnas,Lewenstein}.
It has been shown in~\cite{Sanpera, Vidal}, that any quantum state can be given in terms of local pseudo-mixtures, here and in the following denoted by the quasi-probability distribution $P_{\rm Ent}$, that visualizes entanglement.
An interesting consequence is in this context, that local measurements are sufficient to describe any quantum system completely~\cite{Guehne1,Guehne2}, including entanglement.

In the present contribution we reconsider the representation of entangled states by quasi-probabilities, $P_{\rm Ent}$, in terms of separable states.
This representation is compared with the Glauber--Sudarshan representation of nonclassical states by classical ones.
We give an explicit construction scheme for $P_{\rm Ent}$ for arbitrary quantum states, which opens the possibility of reconstructing these quasi-probabilities from experimental data.
However, the quasi-distributions resulting from the bare reconstruction are not unique.
The identification of entanglement would require to inspect an infinite set of quasi-probabilities.
To overcome this problem is the main aim of our paper.
Starting from the problem that $P_{\rm Ent}$ is not unique, we derive an optimization procedure to obtain $P_{\rm Ent, opt}$, representing a quasi-probability of a given state whose negativities are properly minimized. The most important consequence is that a quantum state is entangled, if and only if $P_{\rm Ent, opt}$ has negativities.
The method can be applied to arbitrary quantum states. It identifies if a state is separable or entangled by a single optimized quasi-distribution.

The paper is structured as follows.
In Sec.~\ref{sec:II} we reconsider both the description of nonclassicality by negativities in the Glauber-Sudarshan representation and the corresponding description of entanglement by negativities in the related quasi-probabilities.
The problem of the reconstruction of quasi-probabilities characterizing entanglement from measured data is considered in Sec.~\ref{sec:III}.
In Sec.~\ref{sec:V} we derive an optimization method for the quasi-probabilities, which is based on the separability eigenvalue problem~\cite{SpeVo1}.
We apply this approach to a separable and an entangled quantum state.
A summary and some conclusions are given in Sec.~\ref{sec:VI}.

\section{Quasi-probability representations}\label{sec:II}

Let us start with the well known Glauber--Sudarshan representation of a single-mode quantum state as an example for quasi-probabilities in statistical quantum physics.
A coherent state can be introduced as the eigenstate of the annihilation operator $\hat{a}$ of a harmonic oscillator, $ \hat{a}|{\alpha}\rangle= \alpha|{\alpha}\rangle$ ($\alpha$ is a complex number).
It has been shown in~\cite{Sudar,Glauber}, that any state can be given in the form of the Glauber--Sudarshan representation
\begin{equation}
	\hat{\rho}=\int dP_{\rm GS}(\alpha) |{\alpha}\rangle\langle{\alpha}|,\label{GSdec}
\end{equation}
with $P_{\rm GS}$ being a signed measure, or more precisely a quasi-propability distribution.
By definition a state is nonclassical, if $P_{\rm GS}$ fails to be represented as a classical propability distribution $P_{\rm class}$, see~\cite{Titulaer,Mandel}.
Thus, a nonclassical state $\hat{\rho}$ has negativities in its $P_{\rm GS}$ distribution. The only classical pure state (with a non-negative distribution $P_{\rm GS}$) is the coherent state~\cite{Hillery}.
Thus the implication of Eq.~(\ref{GSdec}) is, that any quantum state $\hat{\rho}$ can be given solely in terms of classical states $|{\alpha}\rangle$.

As mentioned above it has been shown in~\cite{Sanpera, Vidal}, that any quantum state for finite Hilbert spaces can be given as
\begin{align}
\label{eq:rho}
	\hat{\rho}=(1+\mu)\hat{\sigma} -\mu\hat{\sigma}',
\end{align}
with $\hat{\sigma}$ and $\hat{\sigma}'$ being separable states and $\mu\geq0$.
The condition for positivity, $\langle\psi|\hat{\rho}|\psi\rangle\geq0$, leads to
\begin{align}
\forall|\psi\rangle: \langle\psi|\hat{\sigma}|\psi\rangle\geq \frac{\mu}{1+\mu}\langle\psi|\hat{\sigma}'|\psi\rangle.
\end{align}
A state $\hat{\rho}$ is separable, if it can be written as $\hat{\rho}=\hat{\sigma}$.
This means $\mu=0$.
There are two items describing the failure of an entangled state $\hat{\varrho}$ to be a convex combination of separable states.
For the latter it is not allowed to use coefficient with both $(1+\mu)>1$ and $(-\mu)<0$.

As we have shown in \cite{SpeVo2}, only finite systems need to be considered to characterize entanglement completely.
Hence, it is not a restriction, if one assumes a finite, but arbitrary dimensional Hilbert space $\mathcal{H}_A\otimes\mathcal{H}_B$.
Thus, any quantum state can be given as 
\begin{align}
\hat \rho=\sum_k p_k |a_k,b_k\rangle\langle a_k,b_k|,
\end{align} 
with the quasi-probabilities
\begin{align} 
P_{\rm Ent}=(p_k)_k.
\end{align}
The negative quasi-propabilities can be considered to be the key signature of entanglement.

\section{Reconstruction of quasi-probabilities}\label{sec:III}

This leads us to the question, how to obtain the quasi-probability distribution $P_{\rm Ent}$ -- in particular for entangled states.
We will start to consider two quantum systems represented by $\mathcal{H}_A$ and $\mathcal{H}_B$, given by the bases $\{|k\rangle\}_{k=1,2\dots}$ for each system.
A quantum state can be given, for example, as $|a,b\rangle=|a\rangle\otimes|b\rangle$.
The states $|a\rangle$ and $|b\rangle$ describe the system $A$ and $B$, respectively.
The linear structure of quantum theory allows us to superimpose quantum states.
Let us consider the superposition state
\begin{equation}
	|\Phi^{+}\rangle=\frac{1}{\sqrt{2}}(|0,0\rangle+|1,1\rangle).
\end{equation}
Obviously this state is not factorizable.

To show how to describe entanglement with quasi-propabilities, we start with the Bell state $\hat{\varrho}=|\Phi^{+}\rangle\langle\Phi^{+}|$.
The statistical operator $\hat{\varrho}$ corresponding to this state is given as
\begin{align}
	\hat{\varrho}=&|\Phi^{+}\rangle\langle\Phi^{+}| \\ \nonumber=&\frac{1}{2}
	\left(|0,0\rangle\langle0,0|+|1,1\rangle\langle1,1|+|0,0\rangle\langle1,1|+|1,1\rangle\langle0,0|\right).
\end{align}
To rewrite this state in terms of separable states, we need to define for each system the states $|\mathfrak{s}_n\rangle$,
\begin{equation}
	|\mathfrak{s}_n\rangle=\frac{1}{\sqrt{2}}(|0\rangle+i^n|1\rangle) \quad (n=0,1,2,3),\label{esses}
\end{equation}
and for these states we define factorizable statistical operators  $\hat{\mathfrak{S}}_{n}=|\mathfrak{s}_n\rangle\langle\mathfrak{s}_n| \otimes |\mathfrak{s}_n\rangle\langle\mathfrak{s}_n|$.
Now we can obtain a superposition, operator $\hat{\mathfrak{E}}$, of these separable states $\hat{\mathfrak{S}}_{n}$ with the real coefficients $\pm 1$,
\begin{align}
	\hat{\mathfrak{E}}&=\hat{\mathfrak{S}}_{0}-\hat{\mathfrak{S}}_{1}+\hat{\mathfrak{S}}_{2}-\hat{\mathfrak{S}}_{3}\nonumber \\&=|0,0\rangle\langle1,1|+|1,1\rangle\langle0,0|. \label{EqE}
\end{align}
Therefore, we can rewrite the density operator for the Bell state $|\Phi^{+}\rangle$ in terms of separable states,
\begin{widetext}
\begin{align}\label{eq:bell-sep}
	\nonumber\hat \varrho=&|\Phi^{+}\rangle\langle\Phi^{+}|= \frac{1}{2}(|0,0\rangle\langle0,0|+|1,1\rangle\langle1,1|+\hat{\mathfrak{E}})\\
	=&\frac{1}{2}|0,0\rangle\langle0,0|+\frac{1}{2}|1,1\rangle\langle1,1|+\frac{1}{2}|\mathfrak{s}_0,\mathfrak{s}_0\rangle\langle\mathfrak{s}_0,\mathfrak{s}_0|-\frac{1}{2}|\mathfrak{s}_1,\mathfrak{s}_1\rangle\langle\mathfrak{s}_1,\mathfrak{s}_1|+\frac{1}{2}|\mathfrak{s}_2,\mathfrak{s}_2\rangle\langle\mathfrak{s}_2,\mathfrak{s}_2| -\frac{1}{2}|\mathfrak{s}_3,\mathfrak{s}_3\rangle\langle\mathfrak{s}_3,\mathfrak{s}_3|.
\end{align}
\end{widetext}
For the Bell state $|\Phi^{+}\rangle$, from Eq.~\eqref{eq:bell-sep} we see that the quasi-propability distribution $P_{\rm Ent}$ has negativities,
\begin{align}
	P_{\rm Ent}=(p_k)_k=\left(\frac{1}{2},\frac{1}{2},\frac{1}{2},-\frac{1}{2},\frac{1}{2},-\frac{1}{2} \right).
\end{align}
This is a simple example of a quasi-probability representation of an entangled state by separable ones.

Now we will show that the simple example of a Bell state is sufficient to reconstruct any mixed quantum state.
We may rewrite the matrix $\rho_{klmn}$, representing a general quantum state, in terms of separable quantum states $|a_k,b_k\rangle\langle a_k,b_k|$ and a quasi-propability distribution $P_{\rm Ent}=(p_k)_k$.
Let us consider a density operator of a bipartite system,
\begin{align}
	\hat \rho=\sum_{klmn} \rho_{klmn} |k,l\rangle\langle m,n|,
\end{align}
that has been obtained from experimental data.
By solving the eigenvalue problem of the matrix $\hat \rho$, we obtain the spectral decomposition
\begin{align}
\label{eq:spectral}
	\hat{\rho}=\sum_{i} p_{\phi_i} |\phi_i\rangle\langle\phi_i|,
\end{align}
with $p_{\phi_i}\geq0$ and $\sum_i p_{\phi_i}=1$, which is a common procedure.

It is sufficient to show how to write any pure state $|\psi\rangle\langle\psi|$ solely in terms of separable states.
For any pure state one can obtain the Schmidt decomposition~\cite{book1}
\begin{align}
	|\psi\rangle=\sum_{k=1}^{r(\psi)} \lambda_k |e_k,f_k\rangle,\label{SchmDec}
\end{align}
with $\lambda_k>0$, the Schmidt rank $r(\psi)$, and $\{|e_k\rangle\}_k$ and $\{|f_{k'}\rangle\}_{k'}$ being orthonormal in $\mathcal{H}_A$ and $\mathcal{H}_B$, respectively.
Under the assumption $|\psi\rangle\neq|a\rangle\otimes|b\rangle$, we can write
\begin{align}
	\nonumber|\psi\rangle\langle\psi|=&\sum_{k,l} \lambda_k\lambda_l |e_k,f_k\rangle\langle e_l,f_l| \\
	=&\sum_{k} \lambda_k^2 |e_k,f_k\rangle\langle e_k,f_k| \\
	\nonumber&+\sum_{k>l} \lambda_k\lambda_l \left(|e_k,f_k\rangle\langle e_l,f_l|+|e_l,f_l\rangle\langle e_k,f_k|\right),
\end{align}
The first sum, $\sum_{k} \lambda_k^2 |e_k,f_k\rangle\langle e_k,f_k|$, is a separable state.
The second sum depends on the terms (for $k>l$)
\begin{align}
	\nonumber\hat{\mathfrak{E}}_{k,l}=&|e_k,f_k\rangle\langle e_l,f_l|+|e_l,f_l\rangle\langle e_k,f_k|\\ =&|\mathfrak{s}_0^{(k,l)}\rangle\langle\mathfrak{s}_0^{(k,l)}|-|\mathfrak{s}_1^{(k,l)}\rangle\langle\mathfrak{s}_1^{(k,l)}|\\
&+|\mathfrak{s}_2^{(k,l)}\rangle\langle\mathfrak{s}_2^{(k,l)}|
-|\mathfrak{s}_3^{(k,l)}\rangle\langle\mathfrak{s}_3^{(k,l)}|.\nonumber
\end{align}
with $|\mathfrak{s}_n^{(k,l)}\rangle=\frac{1}{\sqrt{2}}\left(|e_k,f_k\rangle+i^n|e_l,f_l\rangle\right)$.
It can be written analogously to $\hat{\mathfrak{E}}$ as a superposition of separable states $|\mathfrak{s}_n^{(k,l)}\rangle$ with positive and negative coefficients.
According to Eq.~(\ref{eq:rho}), we can give an expansion of any pure state in terms of separable states, $|\psi\rangle\langle\psi|=(1+\mu_\psi)\hat \sigma_\psi-\mu_\psi\hat \sigma_\psi'$, with $\mu_{\psi}=2\sum_{k>l} \lambda_k \lambda_l$  and 
\begin{widetext}
\begin{align}
	\hat \sigma_\psi=&\frac{1}{1+\mu_\psi}\left(\sum_{k} \lambda_k^2 |e_k,f_k\rangle\langle e_k,f_k| +\sum_{k>l} \lambda_k\lambda_l \left(|\mathfrak{s}_0^{(k,l)},\mathfrak{s}_0^{(k,l)}\rangle\langle \mathfrak{s}_0^{(k,l)},\mathfrak{s}_0^{(k,l)}|+|\mathfrak{s}_2^{(k,l)},\mathfrak{s}_2^{(k,l)}\rangle\langle \mathfrak{s}_2^{(k,l)},\mathfrak{s}_2^{(k,l)}|\right)\right),\\
	\hat \sigma'_\psi=&\frac{1}{\mu_\psi}\sum_{k>l} \lambda_k\lambda_l \left(|\mathfrak{s}_1^{(k,l)},\mathfrak{s}_1^{(k,l)}\rangle\langle \mathfrak{s}_1^{(k,l)},\mathfrak{s}_1^{(k,l)}|+|\mathfrak{s}_3^{(k,l)},\mathfrak{s}_3^{(k,l)}\rangle\langle \mathfrak{s}_3^{(k,l)},\mathfrak{s}_3^{(k,l)}|\right).
\end{align}
\end{widetext}
By using Eq.~(\ref{eq:spectral}), any mixed quantum state $\hat \rho$ can be written as a convex combination of pure states.  Therefore, we conclude that any state $\hat{\rho}$ can be written as
\begin{align}
	\hat{\rho}=&\sum_i p_{\phi_i}(1+\mu_{\phi_i})\hat \sigma_{\phi_i}-\sum_i p_{\phi_i}\mu_{\phi_i} \hat \sigma'_{\phi_i}
	\\=&\sum_k q_k |a_k,b_k\rangle\langle a_k,b_k|,
\end{align}
solely in terms of separable states.

We have rewritten the matrix $\rho_{klmn}$, that may be reconstructed from experimental data, in terms of the separable quantum states $|a_k,b_k\rangle\langle a_k,b_k|$ and the quasi-propability distribution $P_{\rm Ent}=(q_k)_k$. In principle,
one may obtain the latter by the procedure presented above.
By experimental state reconstruction one gets $\rho_{klmn}$, the spectral decomposition yields $p_{\phi_i}$ and $|\phi_i\rangle$, via Schmidt decomposition one obtains $\lambda_j$ and $|e_j,f_j\rangle$, which finally leads to the quasi-probabilities $q_k$ and $|a_k,b_k\rangle$.

An arbitrary quantum state can be expressed in an integral form,
\begin{align}
	\hat{\rho}=\int dP_{\rm Ent}(a,b)|a,b\rangle\langle a,b|,\label{eq:intform}
\end{align}
by the non-orthogonal states $|a,b\rangle\langle a,b|$ and the quasi-propabilities distribution $P_{\rm Ent}$ characterizing entanglement.
Negative values of $P_{\rm Ent}$ are necessary signatures of entanglement.
The form of $P_{\rm Ent}$ -- as a signed measure -- can be written by the reconstruction scheme given above as
\begin{align}
	P_{\rm Ent}=\sum_k q_k \delta_{|a_k,b_k\rangle},
\end{align}
with the Dirac measure $\delta_{|a_k,b_k\rangle}$.

\begin{center}
\begin{figure}[ht]
\includegraphics*[width=7cm]{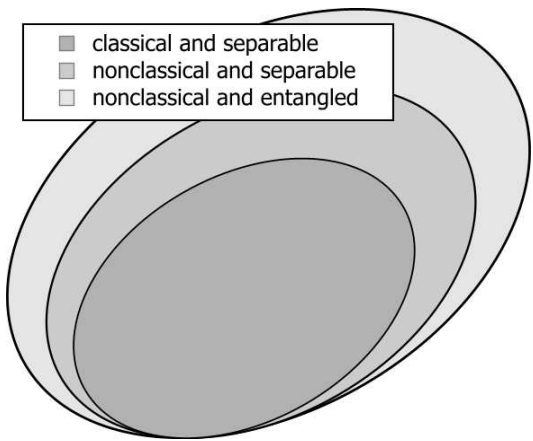}
\caption{Relation between entanglement and nonclassicality of quantum states. A negative $P_{\rm GS}$ quasi-distribution is necessary for entanglement (light gray area). A classical $P_{\rm GS}$ distribution is sufficient for separability (dark gray area). There are quantum states which are separable but not classical ones, see Eq.~(\ref{eq:2fock}).}\label{fig:quantum}
\end{figure}
\end{center}

The integral form of the state $\hat \rho$ in terms of $P_{\rm Ent}$, see Eq.~(\ref{eq:intform}), is analogous to the Glauber-Sudarshan representation with coherent states, see Eq.~(\ref{GSdec}).
In both cases the failure of a classical interpretation of $P_{\rm Ent}$ and $P_{\rm GS}$ describes the quantumness of a state.
Let us assume we have a system of two harmonic oscillators and $P_{\rm Ent}(a,b)=0$ for all factorizeable states $|a,b\rangle$ which are not coherent ones, $|a,b\rangle \not= |\alpha,\beta\rangle$.
In this case we obtain $P_{\rm Ent}=P_{\rm GS}$, with negativities for entanglement.
Furthermore it is obvious that negativities of $P_{\rm GS}$ are not necessarily due to entanglement.
For example, let us consider the Fock state $|n_A\rangle\otimes|n_B\rangle$ ($n_A,n_B\geq 1$),
\begin{align}
	\hat{\rho}=|n_A\rangle\langle n_A|\otimes|n_B\rangle\langle n_B|.\label{eq:2fock}
\end{align}
This is a nonclassical state, but obviously a separable one, $P_{\rm Ent}=P_{\rm class}=\delta_{|n_A\rangle}\delta_{|n_B\rangle}$.
We conclude that entanglement is a subclass of all nonclassical phenomena, see Fig.~\ref{fig:quantum}.

The decomposition of a quantum state as a convex superposition of $p_\phi$ and $|\phi\rangle\langle\phi|$ is not unique.
Thus, the distribution $P_{\rm Ent}$ is also not unique, but at least one reconstruction scheme has been presented in this section.
Let us denote by $\mathcal{N}$ the set of all signed measures $f$, generating the $\hat 0$ operator (for all $|\psi\rangle$: $\hat 0|\psi\rangle=0$),
\begin{align}
	\hat 0=\int df(a,b) |a,b\rangle\langle a,b|.\label{eq:gmeas}
\end{align}
For all $f$ and $f'$ in $\mathcal{N}$ and all real numbers $\nu$ and $\nu'$, the signed measure $\nu f+\nu' f'$ generates again $\hat 0$.
We can write $\hat \rho=\hat \rho+\hat 0$, or we obtain a new measure $P_{\rm Ent}+f$ for any element $f$ of $\mathcal{N}$, that generates the same state.
Under all possible $P_{\rm Ent}$, which generate a given state $\hat \rho$, the quasi-propability $P_{\rm Ent, opt}$ with some kind of minimal amount of negativities needs to be found,
\begin{align}
	\int d(|P_{\rm Ent}(a,b)+g(a,b)|^2)\to \min.\label{eq:minP}
\end{align}
This requires an optimization as presented below in Sec.~\ref{sec:V}.
Obviously, any possible quasi-probability $P_{\rm Ent}$ for an entangled state $\hat \varrho$ has negativities and a state $\hat{\varrho}$ is entangled, if and only if $P_{\rm Ent}$ fails to be a classical propability distribution for any decomposition, $P_{\rm Ent}\neq P_{\rm class}$.
Therefore entanglement is due to negativities in the distribution $P_{\rm Ent,opt}$.

The other way around, as it was known for two decades~\cite{Werner}, any separable state can be given by a classical distribution, $P_{\rm Ent}= P_{\rm class}$.
Thus a single non-negative quasi-distribution $P_{\rm Ent}$ is sufficient to demonstrate the entanglement of a given quantum state.
The advantage of an optimized quasi-probability $P_{\rm Ent, opt}$ is that it clearly uncovers entanglement and separability by its negativities and its non-negativity, respectively.

\section{Optimization}\label{sec:V}
\subsection{Derivation of Optimization}
The Hilbert space $\mathcal{H}_{AB}=\mathcal{H}_A\otimes\mathcal{H}_B=\mathbb{C}^{d_A}\otimes\mathbb{C}^{d_B}$ is finite dimensional, $\dim \mathcal{H}_{AB}=d_Ad_B$.
As we have mentioned above, the assumption that our Hilbert space is finite dimensional is not a restriction for the property of entanglement~\cite{SpeVo2}.
The Banach space ${\rm Herm}(\mathcal{H}_{AB})$ of Hermitian Operators for the given Hilbert space is a finite dimensional real vector space, $\dim {\rm Herm}(\mathcal{H}_{AB})=d_Ad_B$.
First of all we want to introduce the separability norm $\|\cdot\|_{\rm sep}$,
\begin{align}
	\|\hat A\|_{\rm sep}=\sup\{|\langle a,b|\hat A|a,b \rangle|:\langle a,b| a,b\rangle=1\}.\label{eq:defnorm}
\end{align}
Now we prove that the definition given in Eq.~(\ref{eq:defnorm}) is a norm for ${\rm Herm}(\mathcal{H}_{AB})$.
Obviously $\|\lambda\hat A\|_{\rm sep}=|\lambda| \|\hat A\|_{\rm sep}$ and $\|\hat A+\hat B\|_{\rm sep}\leq\|\hat A\|_{\rm sep}+\|\hat B\|_{\rm sep}$ hold true,
\begin{align*}
	\langle a,b|\lambda \hat A+\kappa\hat B|a,b \rangle&=\lambda\langle a,b| \hat A|a,b \rangle+\kappa\langle a,b|\hat B|a,b \rangle\\
	|\langle a,b|\lambda \hat A+\kappa\hat B|a,b \rangle|&\leq|\lambda||\langle a,b| \hat A|a,b \rangle|+|\kappa||\langle a,b|\hat B|a,b \rangle|\\
	\|\lambda \hat A+\kappa\hat B\|_{\rm sep}&\leq|\lambda|\|\hat A\|_{\rm sep}+|\kappa|\|\hat B\|_{\rm sep}.
\end{align*}
Further on $\|\hat A\|_{\rm sep}=0$ is equivalent to $\hat A=0$,
\begin{align*}
	&\|\hat A\|_{\rm sep}=0 \ \Leftrightarrow \ \forall|a,b\rangle:\langle a,b|\hat A|a,b \rangle=0\\
	&\Leftrightarrow \ \forall P_{\rm Ent}: \int dP_{\rm Ent}(a,b)\langle a,b|\hat A|a,b \rangle=0\\
	&\Leftrightarrow \ \forall |\psi\rangle: \langle \psi|\hat A|\psi \rangle=0 \ \Leftrightarrow \ \hat A=0. \quad \blacksquare
\end{align*}

Let $\mathcal{M}$ be the set of separable quantum states.
This set is bounded and closed.
The set $\mathcal{M}_0=\{|a,b\rangle\langle a,b|:\langle a,b| a,b\rangle=1\}$ is the set of pure separable states.
The set of all separable quantum states can be defined as the set of all possible convex combinations of pure separable states, $\mathcal{M}={\rm convex}(\mathcal{M}_0)$.

A quantum state $\hat \rho$ is separable, if and only if it exists a separable quantum state $\hat \sigma$ such that the distance of $\hat \rho$ and this state is zero,
\begin{align}
	&\exists\hat \sigma=\sum_{k}p_k|x_k,y_k\rangle\langle x_k,y_k|: \nonumber\\
	&p_k\geq0, \ \|\hat \rho-\hat \sigma\|_{\rm sep}=0.\label{eq:AAA}
\end{align}
The norm $\|\cdot\|_{\rm sep}$ is the least upper bound ($\sup$) of the projections $\langle a,b|\cdot|a,b\rangle$.
Thus, Eq.~(\ref{eq:AAA}) is equivalent to the optimization
\begin{align}
	g(a,b)=\langle a,b |\hat \rho-\hat \sigma|a,b\rangle\to g_{\rm opt}\label{eq:BBB}
\end{align}
under the condition $h(a,b)=\langle a,b |a,b\rangle\equiv 1$ and the assumption that all optimal (especially the largest and the smallest) values are $g_{opt}=0$.
We discussed an optimization procedure as it is given in Eq.~(\ref{eq:BBB}) in~\cite{SpeVo1}.
For this purpose we derived a set of equations -- the separability eigenvalue equations -- for a linear operator $\hat L$,
\begin{align}
	\hat L_{b}|a\rangle&=g|a\rangle\nonumber\\
	\hat L_{a}|b\rangle&=g|b\rangle,\label{eq:sepev}
\end{align}
with $\langle a,b|a,b \rangle=1$, $\hat L_{b}={\rm tr}_B(\hat L\left[\hat 1_A\otimes|b\rangle\langle b|\right])$ and $\hat L_{a}={\rm tr}_A(\hat L\left[|a\rangle\langle a|\otimes\hat 1_B\right])$.

Obviously the norm is zero, $\|\hat \rho-\hat \sigma\|_{\rm sep}=0$, if and only if all solutions $(g_n,|a_n,b_n\rangle)$ of the separability eigenvalue problem for $\hat \rho$ are solutions for $\hat \sigma$ and the other way around,
\begin{align}
	\hat \rho_{b_n}|a_n\rangle&=g_n|a_n\rangle\nonumber \\ \hat \rho_{a_n}|b_n\rangle&=g_n|b_n\rangle
	\\\Leftrightarrow\nonumber\\
	\hat \sigma_{b_n}|a_n\rangle&=g_n|a_n\rangle\nonumber \\ \hat \sigma_{a_n}|b_n\rangle&=g_n|b_n\rangle.
\end{align}
Thus we have to solve the separability eigenvalue problem of the Hermitian operator $\hat \rho=\hat \sigma$.
We obtain the following equations:
\begin{align}
	g_n=\langle a_n,b_n|\hat \sigma |a_n,b_n\rangle=\sum_k p_k \underbrace{|\langle a_n,b_n|x_k,y_k\rangle|^2}_{G_{n,k}},
\end{align}
with $\vec g=(g_n)_n$, $\vec p=(p_k)_k$ and $G=(G_{n,k})_{n,k}$ this is equivalent to
\begin{align}
	\vec g=G\vec p.\label{eq:opt1}
\end{align}

At least one solution $\vec p_0=(p_k)_k\cong P_{\rm Ent}$ must exist for any $\hat \rho$, since any state can be written in terms of separable states as presented in Sec.~\ref{sec:III}.
The state $\hat \rho$ is separable, if and only if
\begin{align}
	\exists \vec p\mbox{ solution of Eq.~(\ref{eq:opt1})}: (\vec p)_n=p_n\geq0 \ \forall n.
\end{align}
The kernel $\Gamma=\ker(G)$ gives us the possibility to obtain such a positive vector.
It is related to the measure $f$ given in Eq~(\ref{eq:gmeas}), $f\in\mathcal{N}$.
Let $\vec \gamma\in\Gamma$ be an arbitrary element of the kernel.
Thus $\vec p_0+\vec \gamma$ is another solution of Eq.~(\ref{eq:opt1}).
If we rewrite this in therms of measures we obtain $f\cong\vec \gamma$.
The quantum state $\hat \sigma$ is separable, if and only if $(\vec p_0+\vec \gamma)_n\geq0$ for all $n$ for one choice for $\vec \gamma$.

The question arises how to obtain the matrix elements $G_{n,k}=|\langle a_n,b_n|x_k,y_k\rangle|^2$ or what are the states $|x_k,y_k\rangle$.
If we would know the linear map $G$ we could identify its kernel $\Gamma$ and all solutions $\vec p$.
This could solve the separability problem for a given quantum state $\hat \rho$.

For pure separable quantum states $\hat\sigma=|x,y\rangle\langle x,y|$ the only non-trivial solution of the separability eigenvalue problem is $(1,|x,y\rangle)$.
Thus $|x_k,y_k\rangle=|x,y\rangle$ for $g_k=1$ fullfills the desired conditions.
Ignoring this trivial case, let us assume $\hat \sigma$ is a mixed and separable quantum state ($\hat \sigma\notin\mathcal{M}_0$).

As we have mentioned above, the set of separable quantum states $\mathcal{M}$ is a convex set.
Thus it is a subset of a hyperplane of ${\rm Herm}(\mathcal{H}_{AB})$.
Let us shift this convex set $\mathcal{M}'=\mathcal{M}-\hat \sigma$.
Thus the state under study is $\hat{\sigma}'=0$.
The subspace of ${\rm Herm}(\mathcal{H}_{AB})$ given by $\mathcal{V}'={\rm span}(\mathcal{M}')$ and a norm restricted to this space, $\|\cdot\|_{\mathcal{V}'}$, is a normed vector space.
The boundary of $\mathcal{M}'$ with respect to the vector space $\mathcal{V}'$ is $\partial \mathcal{M}'=\mathcal{M}_0-\hat \sigma$.
Thus $\mathcal{M}'$ has a finite volume not equal to zero, $\mathcal{M}'\backslash\partial \mathcal{M}'\neq\emptyset$.

\begin{center}
\begin{figure}[ht]
\includegraphics*[width=8cm]{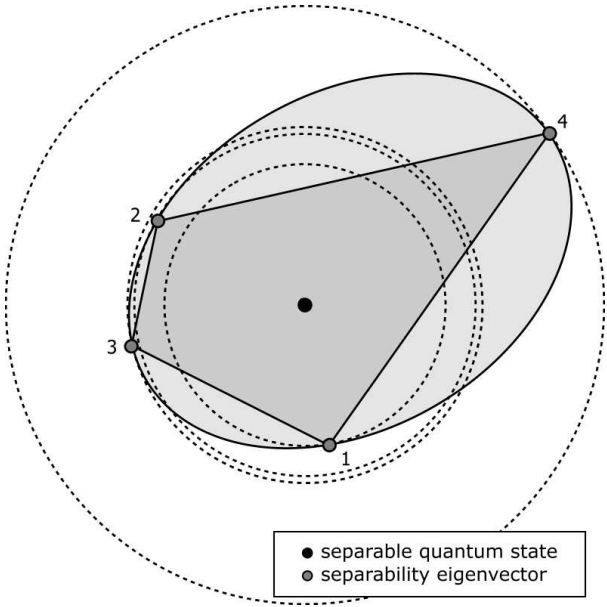}
\caption{The light gray area is the set of separable quantum states $\mathcal{M}$. The radii of the dashed circles are the separability eigenvalues. The dark gray area is the convex set given by the separability eigenvectors. The separable state is element of this set.}\label{fig:convexset}
\end{figure}
\end{center}

Now it is usefull to find the extremal elements of the convex set $\partial\mathcal{M}'$ for the shifted state under study, $\hat \sigma'=0$,
\begin{align}
	\|\hat \sigma'_{\rm pure}\|_{\mathcal{V}'}\to \mbox{optimum} \quad (\sigma'_{\rm pure}\in\partial\mathcal{M}').
\end{align}
Let us denote this optimal elements by $\hat \sigma'_{{\rm pure},i}$.
Since the $\mathcal{M}'$ has a non-zero and finite volume, there are elements $i_k$ such that all $\hat \sigma'_{{\rm pure},i_k}$ are linearly independent (for example in Fig.~\ref{fig:convexset} the vectors from the state to the gray dots 1 and 3), and other elements $i_l$ such that $\hat \sigma'_{{\rm pure},i_l}$ is linearly independent (in Fig.~\ref{fig:convexset} the vectors from the state to the gray dots 2 and 4).
Further on these elements for $i_k$ and $i_l$ give boundaries for $\mathcal{M}'$ in all needed directions.
This means that the tangential hyperplanes of $\partial \mathcal{M}'$ at the points given by $i_k$ and $i_l$ describe a closed set around $\mathcal{M}'$.
Since all $\hat \sigma'_{{\rm pure},i_k}$ and all $\hat \sigma'_{{\rm pure},i_l}$ are directed in this way, the point $\hat \sigma'=0$ can be given as a convex combination of all $\hat \sigma'_{{\rm pure},i_k}$ and $\hat \sigma'_{{\rm pure},i_l}$, 
\begin{align}
	0\in{\rm convex}(\{\hat \sigma'_{{\rm pure},i_k}\}_k\cup\{\hat \sigma'_{{\rm pure},i_l}\}_l).
\end{align}
With other word we can say, that all extremal points of $\hat \sigma'$ at the boundary $\partial \mathcal{M}'$ generate a convex set ${\rm convex}(\{\hat \sigma'_{{\rm pure},i}\}_i)$ which includes this element $\hat \sigma'$, see Fig.~\ref{fig:convexset}.

Since ${\rm Herm}(\mathcal{H}_{AB})$ is finite dimensional, it follows that the norm $\|\cdot\|_{\rm sep}$ is equivalent to any other norm, for example given by $\|\cdot\|_{\mathcal{V}'}$.
Thus the optimization can be performed for the separability norm, which leads to the separability eigenvalues.
Now we can undo the shifting.
Therefore the solutions of the separability eigenvalue problem for $\hat \sigma$ yields our desired vectors $|a_k,b_k\rangle=|x_k,y_k\rangle$.
Thus our map $G$ is given by
\begin{align}
	G_{k,l}&=|\langle a_k,b_k|a_l,b_l\rangle|^2\\\nonumber&={\rm tr}(|a_k,b_k\rangle\langle a_k,b_k||a_l,b_l\rangle\langle a_l,b_l|).
\end{align}
The map $G$ is symmetric and positive semi-definite, since for all real vectors $\vec \pi$ holds
\begin{align}
 	\nonumber &\vec \pi^{\rm T}G\vec \pi=\sum_{k,l} \pi_k G_{k,l} \pi_l \\
	\nonumber =&{\rm tr}\left(\left[\sum_k \pi_k|a_k,b_k\rangle\langle a_k,b_k|\right]\left[\sum_l \pi_l|a_l,b_l\rangle\langle a_l,b_l|\right]\right)\\
	=&{\rm tr}\left[\sum_k \pi_k|a_k,b_k\rangle\langle a_k,b_k|\right]^2\geq 0.
\end{align}
Therefore all eigenvalues of $G$ are positive or zero, and a spectral decomposition of $G$ with real eigenvectors is possible.

In conclusion we showed that the matrix $G$ is given by the separability eigenvectors of the state $\hat \rho=\hat \sigma$.
The matrix $G$ itself is a positive semidefinite symmetric operator and therefore all eigenvectors and eigenvalues can be calculated with algorithms known from linear algebra.
If our state under study is not a convex combination of pure factorizeable states (inseparable), the procedure fails to deliver a solution with $(\vec p)_n\geq0$ for all $n$.
This is due to the fact, that the state is not element of the convex set of separable states $\mathcal{M}$.

\subsection{The Optimization Procedure}

Now we want to describe the optimization scheme as derived above.
We find $P_{\rm Ent,opt}$ for a given quantum state $\hat \rho$ in the following way.
\paragraph*{1.}
We solve the separability eigenvalue problem of $\hat \rho$,
\begin{align*}
	\hat \rho_{b}|a\rangle&=g|a\rangle\\
	\hat \rho_{a}|b\rangle&=g|b\rangle.
\end{align*}
The solutions are given by $(g_n, |a_n,b_n\rangle)$.
\paragraph*{2.}
The vector $\vec g=(g_n)_n$ and the linear map $G=(|\langle a_k,b_k|a_l,b_l\rangle|^2)_{k,l}$ can be calculated.
Now the (ordinary) eigenvalue problem of $G$ can be solved,
\begin{align*}
	G\vec p_\lambda=\lambda \vec p_\lambda.
\end{align*}
A solution of Eq.~(\ref{eq:opt1}), $\vec g=G\vec p$, can be found.
The resulting quasi-distribution for entanglement is $P_{\rm Ent}\cong(p_n)_n=\vec p$, and
\begin{align*}
	\hat \rho=\sum_n p_n |a_n,b_n\rangle\langle a_n,b_n|.
\end{align*}
\paragraph*{3.}
The orthogonal eigenvectors $\vec p_{0_k}$ for the eigenvalue $\lambda=0$ are the basis for the kernel $\Gamma$.
Thus the optimized $P_{\rm Ent, opt}\cong\vec p_{\rm opt}$ and it can be given as
\begin{align*}
	\vec p_{\rm opt}=\vec p +\sum_k c_k \vec p_{0_k},
\end{align*}
with real coefficients $c_k$,
\begin{align*}
	c_k=-\frac{\vec p_{0_k}^{\rm T}\vec p}{\vec p_{0_k}^{\rm T}\vec p_{0_k}}.
\end{align*}
The optimized quasi-distribution is $P_{\rm Ent,opt}\cong\vec p_{\rm opt}$.

Let us prove the given form for $c_k$.
According to Eq.~(\ref{eq:minP}), the minimal amount of negativities needs to be found,
\begin{align*}
	\sum_n |(\vec p_{\rm opt})_n|^2\to \min.
\end{align*}
This is equivalent to the optimization for $(c_k)_k$,
\begin{align*}
	C((c_k)_k)=&\sum_n (\vec p_{\rm opt})_n^2=\vec p_{\rm opt}^{\rm T}\vec p_{\rm opt}\\
 	=&\left(\vec p +\sum_k c_k \vec p_{0_k}\right)^{\rm T}\left(\vec p +\sum_k c_k \vec p_{0_k}\right)\to \min.
\end{align*}
The optimum can be calculated for each $c_l$ as
\begin{align*}
	\frac{\partial C}{\partial c_l}=&2 \vec p_{0_l}^{\rm T}\left(\vec p +\sum_k c_k \vec p_{0_k}\right)\\
	=&2\left(\vec p_{0_l}^{\rm T}\vec p+c_l \vec p_{0_l}^{\rm T}\vec p_{0_l}\right)=0\\
	\Leftrightarrow \ c_l=&-\frac{\vec p_{0_l}^{\rm T}\vec p}{\vec p_{0_l}^{\rm T}\vec p_{0_l}}. \quad \blacksquare
\end{align*}

\subsection{Examples for the Optimization}

To illustrate the strength of our method, we consider two examples: a mixed separable state and a Bell state.
For obtaining analytical solutions, we choose simple examples.
In General, the given optimization procedure can be implemented with numerical methods.
However, the examples provide some basic properties of the optimization.

\paragraph*{Example A.} The first state $\hat \sigma$ is separable, it is a convex combination of five pure separable states,
\begin{align}
	\nonumber\hat\sigma=&\frac{1}{8}|0,0\rangle\langle 0,0|+\frac{1}{8}|1,0\rangle\langle 1,0|+\frac{1}{8}|0,1\rangle\langle 0,1|\\&+\frac{1}{8}|1,1\rangle\langle 1,1|+\frac{1}{2}|\mathfrak{s}_0,\mathfrak{s}_0\rangle\langle \mathfrak{s}_0,\mathfrak{s}_0|,
\end{align}
with $|\mathfrak{s}_0\rangle$ as defined in Eq.~(\ref{esses}).
To solve the separability eigenvalue problem of $\hat \sigma$, we use the Proposition~1 given in~\cite{SpeVo1},
\begin{align}
	\hat \sigma|a,b\rangle=\frac{1}{8}|a,b\rangle+\frac{1}{4}(a_0+a_1)(b_0+b_1)|\mathfrak{s}_0,\mathfrak{s}_0\rangle,
\end{align}
with $|a\rangle=a_0|0\rangle+a_1|1\rangle$ and $|b\rangle=b_0|0\rangle+b_1|1\rangle$.
The vector $|a,b\rangle$ is a separability eigenvector, if $\hat \sigma|a,b\rangle$ has a Schmidt decomposition with $|a,b\rangle$.
We can conclude that the solutions $(g,|a,b\rangle)$ are
\begin{align}
	\left(\frac{5}{8},|\mathfrak{s}_0,\mathfrak{s}_0\rangle\right),
	\left(\frac{1}{8},|\mathfrak{s}_2,\mathfrak{s}_0\rangle\right),
	\left(\frac{1}{8},|\mathfrak{s}_0,\mathfrak{s}_2\rangle\right),
	\left(\frac{1}{8},|\mathfrak{s}_2,\mathfrak{s}_2\rangle\right).\nonumber
\end{align}
Now we obtain $\vec g$, $G$ and the solution of Eq.~(\ref{eq:opt1}),
\begin{align}
	\nonumber\vec g&=\frac{1}{8}(5,1,1,1)^{\rm T}\\
	G&=\left(\begin{array}{cccc}
	1&0&0&0\\0&1&0&0\\0&0&1&0\\0&0&0&1
	\end{array}
	\right)\\
	\nonumber\vec p&=\frac{1}{8}(5,1,1,1)^{\rm T}\cong P_{\rm ent,opt}.
\end{align}
The optimization map $G$ has no eigenvalue equal to zero.
Thus an optimization becomes superfluous.
We conclude that the optimal decomposition of $\hat \sigma$ consists of only four separable pure states,
\begin{align}
	\nonumber\hat\sigma=&
	\frac{5}{8}|\mathfrak{s}_0,\mathfrak{s}_0\rangle\langle\mathfrak{s}_0,\mathfrak{s}_0|+ \frac{1}{8}|\mathfrak{s}_2,\mathfrak{s}_0\rangle\langle\mathfrak{s}_2,\mathfrak{s}_0|\\&+ \frac{1}{8}|\mathfrak{s}_0,\mathfrak{s}_2\rangle\langle\mathfrak{s}_0,\mathfrak{s}_2|+ \frac{1}{8}|\mathfrak{s}_2,\mathfrak{s}_2\rangle\langle\mathfrak{s}_2,\mathfrak{s}_2|.
\end{align}
This example also shows, that the initial pure separable states and their coefficients need not necessarily be the same as the optimization procedure delivers.

\paragraph*{Example B.}
After this simple example let us consider a more sophisticated one.
Therefore let us consider the previously studied Bell state $\hat{\varrho}=|\Phi^{+}\rangle\langle\Phi^{+}|$.
Its importance is due to the reconstruction as presented in Sec.~\ref{sec:III}, which depends only on the reconstruction of this Bell state.

For the solution of the separability eigenvalue problem of $\hat{\varrho}$, let us perform the same procedure as presented in the first example,
\begin{align}
	\hat \varrho|a,b\rangle=\frac{1}{2}(a_0b_0+a_1b_1)(|0,0\rangle+|1,1\rangle).
\end{align}
We obtain the non-trivial solutions,
\begin{align}\label{eq:condbell2}
	\left(\frac{1}{2},|a\rangle\otimes |a^\ast\rangle\right), \mbox{ for } |a^\ast\rangle=a_0^\ast|0\rangle+a_1^\ast|1\rangle
\end{align}
and the trivial ones satisfy
\begin{align}
	a_0b_0+a_1b_1=0.\label{eq:condbell}
\end{align}

If the state is separable, $P_{\rm Ent,opt}(a,b)=P_{\rm class}(a,b)\geq0$, then we can conclude that all needed separability eigenvectors are orthogonal to all $|a,b\rangle$ satisfying Eq.~(\ref{eq:condbell}).
For the example $|a,b\rangle=|0,1\rangle$ we obtain
\begin{align}
	\nonumber\langle 0,1|\hat \varrho|0,1\rangle=&\int dP_{\rm class}(a) \langle 0,1|\left(|a,a^\ast\rangle\langle a,a^\ast|\right)|0,1\rangle
	\\=&\int dP_{\rm class}(a) |\langle 0|a\rangle|^2 |\langle a^\ast|1\rangle|^2=0.
\end{align}
This implies $P_{\rm class}(a)=0$ or $\langle 0|a\rangle=0$ or $\langle 1|a^\ast\rangle=0$.
This means the only needed separability eigenvectors are $|0,0\rangle$ and $|1,1\rangle$, but
\begin{align}
	\hat \varrho\neq \hat\sigma=\frac{1}{2}|0,0\rangle\langle 0,0|+\frac{1}{2}|1,1\rangle\langle 1,1|.
\end{align}
Already at this point, we conclude this state is obviously entangled.

Let us consider the remaining operator $\hat R=\hat\varrho-\hat\sigma$,
\begin{align}
	\hat R=\frac{1}{2}\left(|0,0\rangle\langle 1,1|+|1,1\rangle\langle 0,0| \right).
\end{align}
Obviously it holds $\hat{\mathfrak{E}}=2\hat R$, see Eq.~(\ref{EqE}).
Again the solutions of the separability eigenvalue problem of $\hat R$ are
\begin{align*}
	\left(\pm\frac{1}{4},\frac{1}{\sqrt{2}}(|0\rangle+\exp(i\phi)|1\rangle)\otimes\frac{1}{\sqrt{2}}(|0\rangle\pm\exp(-i\phi)|1\rangle) \right).
\end{align*}
These are states which fulfill Eq.~(\ref{eq:condbell2}) or Eq.~(\ref{eq:condbell}).
Since only 2 independent bases are needed, let us choose $\exp(i\phi)=1,i,-1,-i$.
We obtain states $|\mathfrak{s}_m,\mathfrak{s}_n\rangle$ together with the following order,
\begin{align*}
	\mbox{for $g=+\frac{1}{4}$: }&
	|\mathfrak{s}_0,\mathfrak{s}_0\rangle,|\mathfrak{s}_1,\mathfrak{s}_3\rangle,|\mathfrak{s}_2,\mathfrak{s}_2\rangle,|\mathfrak{s}_3,\mathfrak{s}_1\rangle,\\
	\mbox{for $g=-\frac{1}{4}$: }&
	|\mathfrak{s}_0,\mathfrak{s}_2\rangle,|\mathfrak{s}_1,\mathfrak{s}_1\rangle,|\mathfrak{s}_2,\mathfrak{s}_0\rangle,|\mathfrak{s}_3,\mathfrak{s}_3\rangle.
\end{align*}

The resulting problem is given by $G$, $\vec g$,
\begin{align}
	\vec g&=\frac{1}{4}\left( 1,1,1,1,-1,-1,-1,-1\right)^{\rm T}\\
	G&=\frac{1}{4}\left(\begin{array}{cccccccc} 4&1&0&1&0&1&0&1\\1&4&1&0&1&0&1&0\\0&1&4&1&0&1&0&1\\1&0&1&4&1&0&1&0\\0&1&0&1&4&1&0&1\\1&0&1&0&1&4&1&0\\0&1&0&1&0&1&4&1\\1&0&1&0&1&0&1&4 \end{array}\right).
\end{align}
The eigenvalues $\lambda$ and eigenvectors $p_\lambda$ are
\begin{align*}
	\mbox{for $\lambda=2$: } &\vec p_2=\left( 1,1,1,1,1,1,1,1\right)^{\rm T}\\
	\mbox{for $\lambda=0$: } &\vec p_0=\left( 1,-1,1,-1,1,-1,1,-1\right)^{\rm T}\\
	\mbox{for $\lambda=1$: } &\vec p_{1,1}=\left( 1,0,1,0,-1,0,-1,0\right)^{\rm T}\\&\vec p_{1,2}=\left( 0,1,0,1,0,-1,0,-1\right)^{\rm T}\\
	&\vec p_{1,3}=\left( 1,0,-1,0,0,0,0,0\right)^{\rm T}\\&\vec p_{1,4}=\left( 0,1,0,-1,0,0,0,0\right)^{\rm T}\\&\vec p_{1,5}=\left( 0,0,0,0,1,0,-1,0\right)^{\rm T}\\&\vec p_{1,6}=\left( 0,0,0,0,0,1,0,-1\right)^{\rm T}.
\end{align*}
The vector of separability eigenvalues can be expanded as $\vec g=\frac{1}{4}\vec p_{1,1}+\frac{1}{4}\vec p_{1,2}$.
Thus, the problem $G\vec p=\vec g$ can be solved for example for
\begin{align}
	\nonumber\vec p=&\frac{1}{4}\vec p_{1,1}+\frac{1}{4}\vec p_{1,2}+\frac{1}{4}\vec p_{0}\\
	=&\frac{1}{2}\left( 1,0,1,0,0,-1,0,-1\right)^{\rm T}.
\end{align}
Therefore the operator $\hat R$ is given as
\begin{align}
	\nonumber 2\hat R=&\hat{\mathfrak{E}}=
	|\mathfrak{s}_0,\mathfrak{s}_0\rangle\langle\mathfrak{s}_0,\mathfrak{s}_0|+|\mathfrak{s}_2,\mathfrak{s}_2\rangle\langle\mathfrak{s}_2,\mathfrak{s}_2|
	\\ & -|\mathfrak{s}_1,\mathfrak{s}_1\rangle\langle\mathfrak{s}_1,\mathfrak{s}_1|-|\mathfrak{s}_3,\mathfrak{s}_3\rangle\langle\mathfrak{s}_3,\mathfrak{s}_3|,
\end{align}
which is the form as derived in Sec.~\ref{sec:III}.
Any additional term proportional to $\vec p_0$ does not decrease the number of vectors needed to generate $\hat A$.
Thus the minimal number of vectors needed to generate the state $\hat \varrho$ -- together with $|0,0\rangle$ and  $|1,1\rangle$ -- is six.

Now we can calculate the optimized solution as
\begin{align}
	\nonumber\vec p_{\rm opt}=&\vec p-\frac{\vec p_0^{\rm T} \vec p}{\vec p_0^{\rm T} \vec p_0}\vec p_0=\frac{1}{4}\vec p_{1,1}+\frac{1}{4}\vec p_{1,2}\\=&\frac{1}{4}\left( 1,1,1,1,-1,-1,-1,-1\right).
\end{align}
Thus an optimal decomposition of the state $\hat{\varrho}=|\Phi^{+}\rangle\langle\Phi^{+}|$ in terms of separable quantum states is
\begin{widetext}
\begin{align}
	\nonumber\hat \varrho=&|\Phi^{+}\rangle\langle\Phi^{+}|\\=&\frac{1}{2}\left(|0,0\rangle\langle 0,0|+|0,0\rangle\langle 0,0|\right)\nonumber\\&+ \frac{1}{4}\left( |\mathfrak{s}_0,\mathfrak{s}_0\rangle\langle\mathfrak{s}_0,\mathfrak{s}_0|+|\mathfrak{s}_1,\mathfrak{s}_3\rangle\langle\mathfrak{s}_1,\mathfrak{s}_3|+|\mathfrak{s}_2,\mathfrak{s}_2\rangle\langle\mathfrak{s}_2,\mathfrak{s}_2|+|\mathfrak{s}_3,\mathfrak{s}_1\rangle\langle\mathfrak{s}_3,\mathfrak{s}_1|\right) \\ &-\frac{1}{4}\left(
	|\mathfrak{s}_0,\mathfrak{s}_2\rangle\langle\mathfrak{s}_0,\mathfrak{s}_2|+|\mathfrak{s}_1,\mathfrak{s}_1\rangle\langle\mathfrak{s}_1,\mathfrak{s}_1|+|\mathfrak{s}_2,\mathfrak{s}_0\rangle\langle\mathfrak{s}_2,\mathfrak{s}_0|+|\mathfrak{s}_3,\mathfrak{s}_3\rangle\langle\mathfrak{s}_3,\mathfrak{s}_3| \right).\nonumber
\end{align}
\end{widetext}
This optimal decomposition of the Bell state in terms of separable ones contains negativities.
These negativities are directly related to the entanglement of the state under study.
It differs from the first decomposition given in Eq.~(\ref{eq:bell-sep}).

\section{Summary and Conclusions}\label{sec:VI}

Starting with the fundamental superposition principle of quantum theory, we have demonstrated its equivalence to the concept of quasi-probabilities.
The representation of entangled states by quasi-probabilities, $P_{\rm Ent}$, and separable quantum states has been considered and compared with the Glauber-Sudarshan representation of nonclassical states in terms of classical (coherent) ones.
We have discussed aspects of negativities in the quasi-distributions $P_{\rm Ent}$ in relation to nonclassical Glauber--Sudarshan distributions.
Especially, we have explained how entanglement is embedded into the class of all nonclassical phenomena for the case of a two-mode harmonic oscillator.

Based on the general structure of composite quantum states, we have studied the reconstruction of the quasi-probability distributions $P_{\rm Ent}$ of any quantum state from experimental data.
Starting from a Bell state, we have shown that for all further considerations only this example needs to be understood.
The reconstruction procedure is based on the spectral decomposition of the statistical operator and the Schmidt decomposition of each eigenvector.
Entangled states necessarily have quasi-propability distributions with negativities, independent of the decomposition of the state.
The representation of a quantum state in terms of separable ones and quasi-distributions is not unique.
A positive $P_{\rm Ent}$ proves separability, but negativities are insufficient to verify entanglement.

To obtain a unique description by quasi-probabilities, we have introduced an optimization procedure which delivers an optimal quasi-distribution $P_{\rm Ent, opt}$.
The idea is based on a minimal amount of negativities in the resulting quasi-distribution.
We have derived our optimization on the basis of the recently introduced separability eigenvalue equations.
Negativities in the resulting quasi-probability, $P_{\rm Ent, opt}$, are sufficient to verify entanglement.
The optimization delivers a positive $P_{\rm Ent, opt}$ for separable quantum states.
This approach has been applied to two fundamental examples.

\section*{Acknowledgment}

We gratefully acknowledge valuable comments by M. Piani.
This work was supported by Deutsche Forschungsgemeinschaft through SFB 652.

\end{document}